\documentclass[english,reprint,aps,pre,twocolumn,superscriptaddress]{revtex4-2}

\usepackage[T1]{fontenc}
\usepackage{xcolor}
\usepackage{pifont}
\usepackage{amsmath}
\usepackage{stackrel}
\usepackage{graphicx}
\usepackage{caption}

\makeatletter

\newcommand{\lyxmathsym}[1]{\ifmmode\begingroup\def\b@ld{bold}
	\text{\ifx\math@version\b@ld\bfseries\fi#1}\endgroup\else#1\fi}

\usepackage[T1]{fontenc}
\usepackage{hyperref}
\usepackage{soul}
\usepackage[justification=raggedright]{caption}
\usepackage{tabularx}
\usepackage{multirow}
\usepackage{amsmath}

\usepackage{amssymb}
\usepackage{booktabs}
\makeatother

\usepackage{babel}

\begin{document}
\title{
Transitional patterns on a spherical surface: \\ from scars to domain defects of mixed lattices
\\
}

\author{Wenyu Liu}
\affiliation{School of Physics and Key Laboratory of Functional Polymer Materials of Ministry of Education, Nankai University, and Collaborative Innovation Center of Chemical Science and Engineering, Tianjin 300071, China}
\author{Han Xie}
\affiliation{Department of Physics and Astronomy, University of Waterloo, Waterloo, Ontario N2L 3G1, Canada}
\author{Yu Du}
\affiliation{School of Physics and Key Laboratory of Functional Polymer Materials of Ministry of Education, Nankai University, and Collaborative Innovation Center of Chemical Science and Engineering, Tianjin 300071, China}
\author{Baohui Li}
\email[]{baohui@nankai.edu.cn}
\affiliation{School of Physics and Key Laboratory of Functional Polymer Materials of Ministry of Education, Nankai University, and Collaborative Innovation Center of Chemical Science and Engineering, Tianjin 300071, China}
\author{Jeff Z.Y. Chen}
\email[]{jeffchen@uwaterloo.ca}
\affiliation{Department of Physics and Astronomy, University of Waterloo, Waterloo, Ontario N2L 3G1, Canada}
\author{Yao Li}
\email[]{liyao@nankai.edu.cn}
\affiliation{School of Physics and Key Laboratory of Functional Polymer Materials of Ministry of Education, Nankai University, and Collaborative Innovation Center of Chemical Science and Engineering, Tianjin 300071, China}

\date{\today}
\begin{abstract}
The system of mixed 
hexagonal and square lattices on a spherical surface is examined, 
with an emphasis on the 
exploration of the {disclination} patterns that 
form in the square-rich regime. To demonstrate the possible outcomes, the Hertzian potential energy is used as a model 
for pairwise molecular interactions, which is known to support coexistent hexagonal and square lattices.  
Through molecular dynamics simulations, we show that at least four different 
disclination morphologies arise in a square-rich background: triangular defect domains composed of hexagonal lattices arranged {in} a 
cubic formation, bridged cubic state, linear scar {disclinations} with no hexagon content, and open scar disclinations 
containing a significant amount of 
hexagonal lattice in the open regions. 
Order parameters are also introduced to highlight the significance of the bridged and open-scar {disclinations}, both being 
the new morphologies reported in this study. The fact that the bridged state is an energetically preferred one is further demonstrated by a separate elastic energy model, which confirms its prevalence
over the unbridged cubic state. 

\end{abstract}
\maketitle

When particles are densely packed on a spherical surface, what structures can be observed?
On a flat surface, dense packing typically leads to a uniform hexagonal lattice structure, with particles positioned at the lattice sites. In less common cases, a square lattice may also form.
On a spherical surface, 
the lattice textures are not only distorted, but are also accompanied by disclinations, due to the imposed 
geometry constraint. These disclinations are
necessitated by the Poincar{\' e}-Hopf theorem, which dictates the total sum of the disclination winding numbers \cite{poincare1895analysis}.
A well-known example is the Casper-Klug (CK) construction for triangular lattices on a spherical surface, which 
exhibits twelve pentagonal disclinations arranged in icosahedral symmetry on the surface \cite{Caspar1962}. 
This effect,  known as the emergence of 
 topological defects, appears in practical systems under geometric frustrations; some well cited examples are the surface of viral capsids \cite{Caspar1962,Sleytr2014,Laughlin2022},
 monolayer assemblies of colloidal beads on the surfaces of spherical liquid droplets \cite{Irvine2010,Irvine2012,Meng2014,Guerra2018,Das2022}, and liquid crystal shells on the surfaces of spherical water droplets in double emulsions \cite{Lopez-Leon2011,lopez2011nematic2,li2016colloidal}.

 Understanding defect structures on curved surfaces
 helps us better understand the role geometry plays in soft matter and materials science
 \cite{Lubensky1992, Nelson2002,Bowick2009}. 
For particle assemblies on a spherical surface that displays a single type of lattice, hexagonal ({Hex}) or square (Sq), much progress has been made in characterizing their defect patterns, theoretically and experimentally \cite{Bowick2000interacting,Bausch2003,van2006symmetry,Plevka2008,Irvine2010,Irvine2012,Meng2014,Manoharan2015, Guerra2018,Das2022,domonkos2022nanosphere}.
Two types of disclinations are commonly possible:  point defects such as the localized pentagon regions in CK construction \cite{Caspar1962} and 
scar disclinations that form visibly distinct lines on the spherical surface consisting
of local positive and negative disclination pairs
\cite{Bausch2003,giomi2007crystalline,bowick2008bubble,zhu2024programmable,xie2025}. 

Particle assemblies that display the mixing of two types
of lattices introduces another level of complexity, due to the significant involvement of lattice competition. In hard-sphere systems, crystals typically adopt the hexagonal phase to maximizes spatial efficiency. In contrast, systems described by soft-sphere interactions can give rise to other phases \cite{Miller2011}; for instance, square phases have been observed in systems governed by the Hertzian potential \cite{Zu2016,Xu2019}. The coexistent Hex-Sq crystalline–crystalline lattice in flat geometries is experimentally observed and analyzed in metallurgy \cite{Reed1973,Perim2016}, extraordinary ice \cite{Huang2023,Knopf2023}, and soft-core particle assemblies \cite{Frenkel2002,Frenkel2006,Zhao2011,Manoharan2015,Peng2015,Rossi2015,Li2016,Rey2017,Singh2022}. 
On a spherical surface where curvature takes a dominant role, 
we are only beginning to understand the sophisticated defect structures. The most important 
feature is the emergence of area-related defects as a compound structure, beyond point and line defects in a single-phase lattice.

In a recent study, we investigated the defect patterns of coexistent 
Hex-Sq lattices on spherical surfaces using molecular dynamics simulations of a known
molecular system that shows such coexistence on a flat surface \cite{xie2025}.
We found that crystalline–crystalline phase separation on the surface gives rise to domain defects, where one lattice type forms multiple distinct domains embedded within the bulk of the other, resulting in high global
symmetry across the sphere. 
From the Hex side, as the Sq lattice area fraction $f_{\text{Sq}}$ increases from 0, 
regimes of the low-energy morphologies are identified:  
a CK-like or linear-scar structure in the Hex lattices for $f_{\text{Sq}}\sim 0$, 
enclaved Sq domain defects and their Sq-hex compound structures  
due to the Gauss-Bonnet spherical-area requirements as $f_{\text{Sq}}$ increases,  
enclaved Hex domain defects in a mainly Sq background, and finally 
linear-scar defects in a mostly uniform Sq lattice when $f_{\text{Sq}} \sim 1$.

This paper reports
previously unidentified dislocation types, revealed through a more detailed exploration of 
the Sq-rich regime. In the $f\lesssim 1$ regime that $f_{\text{Sq}}$ is close to 1, as the Hex content increases, 
a linear-scar grows into an open-scar, where in
the open area, narrow Hex corridors start to emerge. 
This distinct state, coming from the $f_{\text{Sq}}\sim 1$ side by lowering $f_{\text{Sq}}$, 
starts to host Hex-Sq coexistent line boundaries, which are interrupted and connected by the $\pm 1/12$
disclination pairs. Lowering $f_{\text{Sq}}$ promotes widening of the open area. 
The structure consistently exists in the large $N$ limit, becoming a common feature of the Sq-rich state. 

Another main discovery is the existence of the bridged state. At a first glance, it
can be viewed as 
a modification of the unbridged cubic defect structure, where triangular Hex-domain defects 
reside at the corners of a cube, in the background of Sq lattice. 
The formation of the bridges between these domains continuously 
shifts the locations of the defect points originally associated with the unbridged Hex domains. 
As it turns out, the bridged morphology has a lower system energy, and becomes more prevalent in the large $N$ limit, 
ultimately replacing the unbridged regime in the state diagram. 

To further prove that this is an energetically preferred 
morphology in the continuum limit, a defect-point-to-defect-point 
elastic energy, which effectively incorporates the lattice distortion energy into a point-based framework \cite{Bowick2002},
is employed. This calculation is entirely independent of the MD simulations and provides 
another theoretical perspective. 
Based on a comparison between the elastic energies of the bridged and unbridged states, 
the results are in agreement with the MD findings.

The finding of the bridged state and open-scar substantially broadens the understanding of defect structures on spherical surfaces, revealing novel morphologies that arise uniquely in mixed-lattice systems. These findings highlight the rich diversity of topological defects shaped by curvature, frustration, and lattice incompatibility. Together with Refs. \cite{xie2025,xie2025filling}, the study presented in this paper offers
new insights into curvature-driven disclination 
structures formed on spherical surfaces. A {more} complete physical picture begins to emerge to capture the common
topological properties of 
lattice mixing, despite the diversity of defect morphologies on spherical surfaces.

\begin{figure*}[!t]
\centering
\includegraphics[width=1\textwidth]{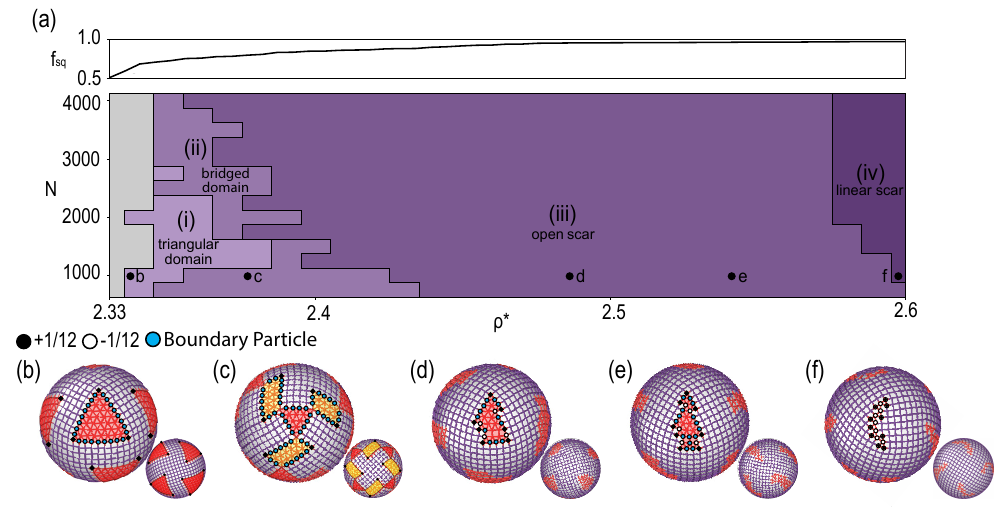} 
\caption{\label{fig:wide-1}State diagram and snapshots. As $\rho^{\ast}$ increases,
four defect modes are observed: (b) $(2.34,1000)$ triangular domain, corresponding to regime-(i); (c) $(2.37,1000)$  {linked domain}bridged domain, regime-(ii);
(d) $(2.49,1000)$ and (e) $(2.55,1000)$  {fat scar}open scar, regime-(iii); and (f) $(2.60,1000)$ linear scar, regime-(iv). (b-f) locations in the state diagram are labeled by solid circles in (a). The Hex lattices and defects are colored red, while Sq lattices purple. The upper panel shows $f_{\text{Sq}}$, the area fraction of Sq lattices for $N=1000$
as a function of particle density $\rho^*$. }
\label{Fig1}
\end{figure*}

\smallskip
\noindent \textbf{Simulation method.} 
A particle-based study is carried out by
conducting Langevin dynamics of $N$ particles confined to a spherical surface of radius $R$, where
the particles interact with each other through the Hertzian potential.
The total potential energy of the system is minimized through a temperature annealing
protocol, implemented using LAMMPS \cite{LAMMPS}. Twenty independent runs
were conducted to select configurations corresponding to low-energy states for analysis purposes. The pairwise Hertzian potential, given in Appendix \ref{App:simulation}, is a repulsive
elastic interaction that vanishes beyond a characteristic force range $\sigma$ \cite{Landau1970,Johnson1985}.
Two dimensionless parameters, $N$ and the
reduced surface density $\rho^{\ast} = N\sigma^2 / 4\pi R^2$, characterize
the system.
The defect states have been analyzed in
recent MD studies 
on flat
\cite{Yao2020}
and spherical surfaces \cite{Miller2011,Yao2020,xie2025}. 

\smallskip
\noindent
\textbf{State diagram.} Several earlier works have extensively investigated Hex-Sq coexistence of Hertzian particles on a plane \cite{Miller2011,Zu2016,Zu2017,tsiok2021structural}, all of which indicated the coexistence density range $2.2\lesssim\rho^*\lesssim2.4$. On a spherical surface, all density we focused in this work appear near these densities.
The defect-state diagram on a sphere for $\rho^* \lesssim 2.33$, where the background
lattice is dominated by triangles and hosts various types of disclination patterns,
has been reported recently \cite{xie2025}. Here, in
Fig. \ref{Fig1}(a), the focus shifts to the regime
of the state diagram that features unique defect morphologies, in a background
rich in Sq lattices, within the density range $2.33 \lesssim \rho^{\ast} $. The state diagram is constructed by extensively sampling selected locations across the entire $(\rho^*, N)$ parameter space shown.

The state diagram is color-shaded to delineate parameter regimes corresponding to distinct defect patterns. 
On the left, in the gray regime, $\rho^*$ falls within the first-order Hex–Sq transition densities observed in flat space, where maze-like, irregular, phase-separated patterns are typical, for which 
no clear, symmetric defect patterns can be identified. 
On the right, as $\rho^*$ increases, 
four {distinct}, ordered patterns emerge within the sea of Sq lattices.
These include 
those with 
(i) eight Hex domains exhibiting cubic symmetry $O$, (ii) Hex
bridges connecting the eight linked Hex domains, (iii) open Hex scars in the 
torn region, and (iv) thin scars composed of mostly paired positive-negative defects.
Typical simulation snapshots are presented in Fig.~\ref{Fig1}(b)-(f), and their location in the parameter space $(\rho^*,N)$ is labeled in Fig.~\ref{Fig1}(a).
At the top of Fig.~\ref{Fig1}(a), the area fraction of Sq lattices $f_{\text{Sq}}$, measured from the simulated configurations, is plotted as an indicator of {the} Sq-lattice content.

The state diagram has been constructed from data sampled at every density with an interval of $0.01$, and at every $N$ with an interval of 250 ($N=750, 1000, 1250, \dots, 4000$). Phase boundaries are identified using structural parameters and confirmed by visual inspection.
The rugged and discontinuous appearance of the phase boundaries mainly arises from two factors. The energy differences between competing states along the boundaries are small. Although 20 independent MD annealing runs were performed, the final determination still carries some uncertainty, contributing to boundary roughness. In addition, near the identified boundaries, different types of defects may coexist. A majority-pattern rule was applied in these cases, which also adds to the apparent ruggedness of the boundaries.

\smallskip
\noindent
\textbf{Linear and open scars.} In the high $\rho^*$ regime, the surface is predominantly
 covered by Sq lattices.
In the ideal case, a defect 
point corresponds to missing Sq cell carrying a  {winding} number of $+1/4$, located at the 
junction where three nearly uniform Sq lattices meet. 
In total, eight 
such points are needed to ensure that the total winding number is  $+2$, meeting the requirement of the spherical geometry \cite{poincare1895analysis}. 

This, however, is not energetically most favorable in a Hex-Sq coexisting 
lattice system where a local 
triangular cell is available. 
As demonstrated in Fig. \ref{app_fig2}, a single triangular cell,  {or a pair of them} replacing a square, creates a
defect point with a winding number of
 $+1/12$. In contrast, the substitution of a square for a triangular cell gives rise to a $-1/12$ defect. Instead of a $+1/4$ Sq defect,  a disclination identity
 composed of 
 multiple such defect points, while 
 maintaining a net winding number of $+1/4$,
can exist. One of such identity is the so-called scar, where a linear chain is formed from 
alternating $(+1/12)$-$(-1/12)$ defect points, embedding three extra $+1/12$ points along the scar
[see the snapshot in Fig. \ref{Fig1}(f)].
This is the core idea behind the scar-disclination theory proposed
in \cite{Bowick2000interacting}, for both Hex and Sq lattices on a spherical surface.  
Scar disclinations in Hex lattice have been observed experimentally 
through the assembly of colloidal particles on a spherical surface 
\cite{Bausch2003}. 
In both Hex and Sq lattices, particle-based computer simulations have also confirmed the existence of scars \cite{xie2025}.

In the lower $\rho^*$ regime, labeled as region (iii) in Fig.~\ref{Fig1}, a higher content of Hex lattices must be present. A linear scar opens into a narrow corridor 
containing Hex lattices. Clear boundaries between Hex and Sq domains start
to emerge and occupy a significant portion of the open-scar morphology. Each boundary particle, marked in cyan in the figure, serves as a junction of two Sq and three triangular cells, and carries a winding number of $+0$. Accompanying the emergence of these boundaries, the number of $\pm 1/12$ defects along the open scars is reduced. Typically, there are eight open scars on the spherical surface, the same as the number of linear scars. Each scar carries
a compound disclination of $w=+1/4$.

\begin{figure}[!ht]
\centering
\includegraphics[width=1\columnwidth]{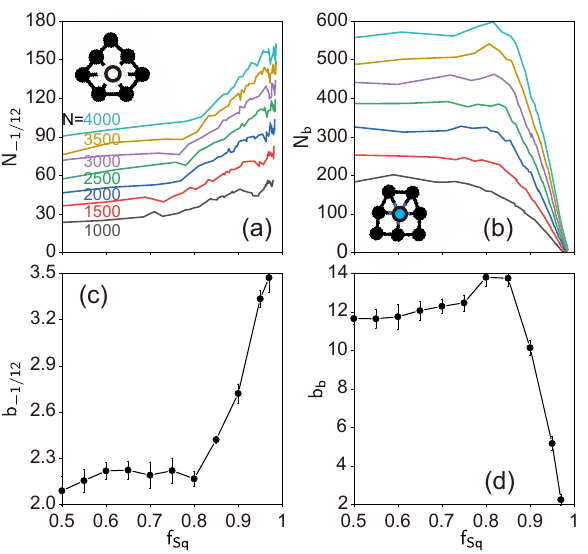} 
\caption{Number of disclination $N_{-1/12}$ and number of Hex-Sq boundary particles $N_b$ as a function of $f_{\text{Sq}}$. Using the ordinary least sqares method, the original data in (a) and (b) are fitted to the forms $N_{-1/12} \sim a_{-1/12} + b_{-1/12} {\sqrt N}$ and $N_b \sim a_b + b_b {\sqrt N}$, respectively,  to extract the asymptotic scaling coefficients $b_{-1/12}$ and $b_b$ at large $N$. The fitted relationships of $N_b$ and $N_{-1/12}$ with $\sqrt{N}$ are presented in Fig. \ref{app_fig2} of Appendix \ref{App:fitting}.
The curves in (a) and (b) share the same color code for the total particle number $N$ in the system: 
$N=1000$ (grey), $1500$ (red), $2000$ (blue), $2500$ (green), $3000$ (purple), $3500$ (orange), and $4000$ (light blue). 
The inset in (a) illustrates a $-1/12$ defect [white particles in Fig.~\ref{Fig1}(d)-(f)], and 
the insert in (b) a Hex-Sq boundary particle [cyan particles in Fig.~\ref{Fig1}(b)-(e)].  
}
\label{FIG2}
\end{figure}

A  power-law scaling 
$N^{1/2}$ for the number of defects involved in 
linear scars was originally {suggested} in \cite{Bowick2000interacting}, 
and numerically verified by computer simulation data \cite{xie2025}.
Here, 
Figure ~\ref{FIG2}(a) and (b) displays $N_{-1/12}$, the number of particles in the center of the
$w=-1/12$ defects, and $N_b$, the number of Hex-Sq boundary particles, as the signatures of the appearance of open scars. Because these
particles scatter along {\emph{linear}} boundaries, 
$N_{-1/12}$ and $N_b$ of open scars also follow the same {power-law} behavior
for a given $f_{\text{Sq}}$, at large $N$ limit. 
We use here
$N_{-1/12} \sim a_{-1/12} + b_{-1/12} {\sqrt N}$ and $N_b \sim a_b + b_b {\sqrt N}$.
Although 
$N$ has been pushed to a large number in the simulation, after taking the square root,
the finite size effects remain, which necessitates the introduction of the $a$ constants.

The plots in Fig.~\ref{FIG2}(c)  and (d) clearly show how open scars close towards linear scars around $f_{\text{Sq}}\sim 0.97$
as the Hex content decreases. Open scars are observed in the approximate range 
$f_{\text{Sq}}\sim (0.8, 0.97)$.
Characteristically, from the $f_{\text{Sq}}\sim 1$ end, 
$N_{-1/12}$ 
declines until $f_{\text{Sq}}\sim 0.8$ but $N_b$ remains approximately constant. Below $f_{\text{Sq}}\sim 0.8$, the state of 
bridged-domain defects takes over.

The coefficients extracted from the ${\sqrt N}$-scaling fits have clear physical meaning:  $b_{-1/12}$ represents the line density of -1/12 disclinations, while $b_b$ is equivalent to the coefficient of the total length (perimeter) of the neutral Hex–Sq interfaces, expressed in units of the lattice spacing. As $f$ decreases in the open-scar regime ($0.97\gtrsim f\gtrsim0.8$), neutral interfaces replace alternating ±1/12 pairs, causing $b_{-1/12}$ to fall, whereas the perimeter coefficient $b_{-1/12}$ increases. Across the subsequent transition to the bridged state ($f\lesssim0.8$), the connectivity of interfaces changes, but their overall length scaling is preserved, so $b_b$ remains essentially unchanged while $b_{-1/12}$ stays low.

In addition, as bridged states transform into open scars, isolated dislocations emerge in the bulk, giving rise to multiple small hexagonal domains. These domains contribute appreciably to the measured value of $N_b$. In contrast, $b_{-1/12}$, shown in Fig.~\ref{FIG2}(c), exhibits no sudden variation associated with these dislocations, indicating that they merely migrate from the boundary into the bulk without altering their total number.

\smallskip

\noindent\textbf{Bridged domain defects.} 

An ideal defect structure consisting of  {eight}  well-formed triangular, Hex-rich 
regions is shown in Fig. \ref{Fig1}(b), where the centers of the Hex domains form
a cubic symmetry on the spherical surface. 
This was designated as Hex-domain defects in a Hex-Sq mixing 
system when $f_{\text{Sq}}$ exceeds $\sim 0.6$ in Ref. \cite{xie2025}. It is a natural extension of 
the  {octahedral} $O$ symmetry formed by defects in a Sq-only lattice, as the defect points in that case already exhibit
triangular geometry, but in localized spherical regions \cite{xie2025filling}.

A careful examination reveals that 
the defect structures in
a significant parameter regime includes {a} previously unreported type, in which bridges form between Hex-rich triangular regions.
This novel defect structure exhibits complex morphology yet is energetically favorable, as demonstrated by 
two separate independent theoretical methods in this study, MD and defect elastic theory. 
Figure. \ref{Fig1}(b) and (c) show MD snapshots of unbridged and bridged states, respectively.
The latter consists of eight primary 
triangular domains in red and the bridges are highlighted in {yellow}.

\begin{figure}[!t]
\centering
\includegraphics[width=1\columnwidth]{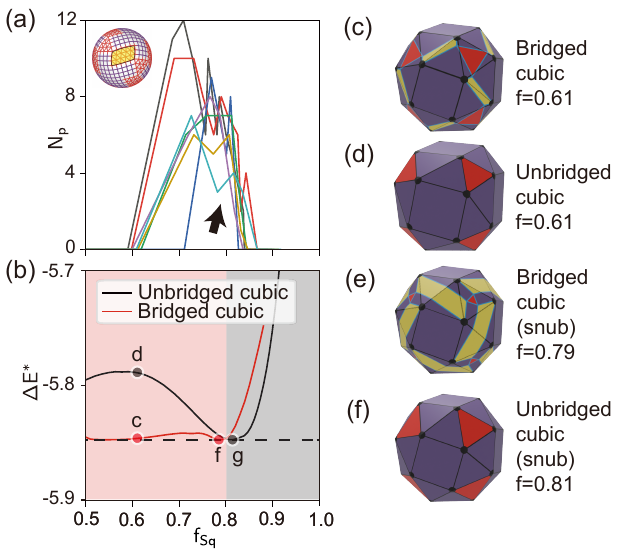} 
\caption{\label{FIG3}Unbridged and bridged cube configurations and their energies. 
In plot (a), the 
number of parallelogram bridges, $N_p$, is plotted as a function of $f_{\text{Sq}}$, where the color represents the same $N$ values used in Fig. \ref{FIG2}.
The inset describes such a parallelogram by  {a} yellow shade. In plot (b) 
the 
minimized BNT elastic energy $\Delta E^*$ (Eq.~\ref{effect_BNT}), for both
states are shown{,} and the bridged state has a lower $\Delta E^*$ in the regime $f_{\text{Sq}}\leq 0.80$
A dashed line is plotted for the ideal snub cube configuration,
 at $\Delta E^*\sim-5.85$, for $n=24$ free particles interacting via the BNT potential on a sphere. 
 In panels (c)-(f), the polyhedral view{s} of the bridged and unbridged states are shown at {the} indicated 
 values of $f_{\text{Sq}}$. Among these, the polyhedrons in (e) and (f) are the snub cube.
 }
\end{figure}

In a typical pattern of unbridged cubic domain defects, shown in Fig. \ref{Fig1}(b), 
the edges of neighboring domains are parallel on a spherical surface. The entire pattern exhibits a 
 cantellated cube symmetry in the polyhedron view \cite{xie2025filling}, shown here in Fig. \ref{FIG3}(d). 
In the ideal bridged state, two main types of Hex domains are present, one forming an overall triangular shape [red in 
Figs. \ref{FIG3}(c) and (e)] and the other a parallelogram-shaped bridge ({yellow}).
The nearest edges of the different red domains remain parallel.  The relative orientations of the 
original triangular Hex domains are rotated about the axis connecting 
the sphere center and the Hex domain center, to accommodate the orientations of the bridges. 
The $w=+1/12$ defect, which originally resides at the triangular corner, now translates to the 
sharp corner of the parallelogram. 
The triangular corner now hosts a junction where two square lattice cells and three triangular lattice cells meet, which is a $w = 0$ point, a location without real disclination.

The number of spherical parallelograms, $N_p$, serves as a distinct order parameter of the bridged state. In an ideal case, there are $N_p=12$ of these and the entire defect pattern on the sphere forms a linked arrangement comprising 
eight triangular domains and twelve parallelogram domains. In MD simulations, a perfectly ordered state rarely exists. Figure~\ref{FIG3}(a) shows the measured values of $N_p$ produced from the simulations, which indicates that the bridged state appears in the approximate range $f_{\text{Sq}}\sim (0.6,0.8)$.

It is instructive to see that the number of $w=0$ junctions associated with  {the} line boundary between the Hex-Sq domains $N_b$, marked by cyan circles in Fig.~\ref{Fig1}(b)-(d), remains constant when the state boundary between bridged state and open-scar state is crossed by changing $f_{\text{Sq}}$. This can be observed in Fig.~\ref{FIG2}(c), where $N_{-1/12}$ in (a), a signature of the open-scar state, declines
when the state boundary is crossed from the open-scar side at $f_{\text{Sq}}\sim 0.8$.

\smallskip
\noindent
\textbf{BNT elastic energy for the bridged state.} 
To understand why the bridged state has a lower system energy than the  cubic state consisting of 
triangular domains, here an elastic energy analysis, which is a different origin from MD simulations and suitable for the continuum limit, is performed. 
In the context of an ``ocean'' of background
lattices, the BNT model, introduced
by Bowick-Nelson-Travesset \cite{Bowick2000interacting}, offers a simplified view
of the system energy. The system energy is effectively written as the sum of defect-point to defect-point interaction potential, with no involvement of the actual background lattice types making up the system. 
For the same types ($w=+1/12$) and number of defects ($n=24$) {for the unbridged and bridged cubic states}, the BNT energy can be further simplified  to {(see details in Appendix} \ref{Appendix A})
 \begin{equation}
    \Delta E=\frac{1}{2} \sum_{i=1}^{n}\sum_{j=1}^{n}\chi(\beta_{ij}).
    \label{effect_BNT}
\end{equation}
 where $\beta_{ij}$ is the geodesic angle between defect points $i$ and $j$. The interaction energy 
 $\chi(\beta)$ is a soft repulsive interaction, expressed and plotted in 
Appendix \ref{Appendix A}; it favors
 configurations in which two interacting defect points are maximally separated in distance.
The analysis requires minimizing $\Delta E$ under the geometric constraints that govern the
arrangements of the domain defects in the unbridged [Figs.~\ref{FIG3}(d) and (f)] and bridged [Figs. \ref{FIG3}(c) and (e)] states, for a given $f_{\text{Sq}}$. 
The degrees of freedom include the collective rotational angle 
of the triangular domains about the spherical axis. The minimized energy is denoted 
as  $\Delta E^*$.
Figure~\ref{FIG3}(b) illustrates the minimized elastic energies of these two defect states.  
It can be clearly seen that the bridged state has lower energy when $f_{\text{Sq}}< 0.8$, consistent 
with the location of the state boundary identified from the MD simulations in Fig.~\ref{FIG3}(a), indicated by an arrow in the {figure}. 

When $n$ BNT-interacting point disclinations are placed on a spherical surface without imposing any 
geometric constraints, the problem becomes analogous to the classical Thomson problem, which seeks the energy-minimizing configuration of $n$ particles, interacting via Coulomb potential, on a spherical surface. 
The solution for $n=24$ free BNT-interacting disclinations yields a snub cube configuration,  
an Archimedean solid composed of 6 square faces and 32 {\emph {equilateral}} triangles in a polyhedral view, on the 
same polyhedron shown in Figs. \ref{FIG3}(e) and (f). For the BNT potential, the snub energy 
 $\Delta E^*$ has a value $-5.847$, indicated by the 
 dashed line in Fig.~\ref{FIG3}(b). 
The defects in the bridged cubic and unbridged cubic states reach the snub configuration at $f_{\text{Sq}}=0.787$ and $0.814$, respectively.

 Generally, within the range of $f_{\text{Sq}}\sim (0.6, 0.8)$, the configurations adopted by both the
 bridged and unbridged states resemble the snub cube configuration, also featuring 
 6 square faces but 32 {\emph {unequilateral}} triangles in a polyhedral view. For the same values of
 $f_{\text{Sq}}$, the additional flexibility in arranging the parallelograms in the bridged state allows the defect points to be positioned closer to the corners of the ideal snub cube configuration. 
Figures~\ref{FIG3}(c) and (d) illustrate an example of $f_{\text{Sq}}=0.61$
bridged and unbridged cubic configurations. It is visually obvious that the bridged state 
more closely approximates the geometry of the ideal snub cube.

Why does the bridged state give way to 
the open-scar state in a higher $f_{\text{Sq}}$?
As the particle density increases, both the bridged and unbridged states rapidly deviate from the snub cube configuration.
Beyond $f_{\text{Sq}}\sim 0.81$, this deviation 
becomes significant enough that the elastic energy produced by defect interactions in these cube-based states grows substantially. 
On the other hand, an open scar configuration can well-place 
 primary defect points, those located at the end of a scar, by a considerable distance, in a Sq lattice dominant state $(f_{\text{Sq}}\lesssim 1)$. The appearance of paired defects, $\pm 1/12$, is a feature that essentially minimizes the local elastic energy, which, in the continuum limit, effectively cancels each other's elastic distortions.  
This defect-pairing feature is absent in the cube-based configurations, giving the open-scar state a distinct energetic advantage at high 
$f_{\text{Sq}}$.

\smallskip
\noindent
\textbf{Summary.} 
This study focuses on Sq-rich lattices confined on a 
spherical surface, where Hex-lattice appears as domain disclinations in the mix. 
The dominant parameters are the system size, $N$, and the particle density, $\rho^{*}$. 
In a system where Sq and Hex lattices can coexist, which is the case for Hertzian soft particles, 
this gives rise to a rich paradigm for the competition between different types of disclinations. Within
the parameter
range examined, unbridged and bridged cubic morphologies, as well as open- and linear-scar configurations, are shown to exist. Among these, the bridged cube and open scar states are described for the first time in the literature. 
The morphological transition from unbridged to bridged states is attributed to   
the 
competition of defect stress energies, from a continuum theory without the involvement of 
the actual, molecular-level interactions. 

This study complements the recent 
analysis  of
defect morphologies appearing in Hex-rich lattices, those with $f_{\text{Sq}}\lesssim 0.5$, as 
 reported in Ref. 
 \cite{xie2025}.
 The overall picture then emerges: 
 when Hex and Sq lattices are in the mix on a spherical surface, the possible types of defect morphologies include domain defects, enclaved
 and double-enclaved domain defects \cite{xie2025},  bridged domain defects, and open-scar defects, on top of previously 
 known point and linear-scare disclinations that occur in unmixed lattices \cite{Bowick2009}. The mixing may arise from the co-existence of Sq and Hex lattice in a molecular system, or from placing  
 weakly deformed Sq and Hex tiles on a spherical surface   \cite{xie2025filling}. 
The identification of the bridged state and open-scar significantly expands our knowledge of defect structures on spherical surfaces, uncovering new morphologies that emerge specifically from mixed-lattice interactions. These results underscore the diversity of topological defects driven by curvature, frustration, and lattice mismatch, providing new perspectives on self-assembly on curved substrates with relevance to soft matter, materials engineering, and biological systems, and offering new ways for materials design. 

Some of {the} states discovered here can be readily verified experimentally by extending flat Sq-Hex lattices  \cite{Lee2008,Pang2013,Zarzar2015} to curved spherical geometries. Building upon our studies, 
mixed lattices on more  {sophisticated}
curved surfaces, such as ellipsoids, toroids, and hyperbolic geometries, where curvature anisotropy arises,
could lead to other novel defect effects. In our studies, 
we assumed  {a} single-layer molecular film on a spherical surface; mounting multiple layers of molecules can give rise to new physics, as shown by experiments \cite{Peng2015}, which, on a spherical surface where topological defects are 
a  {necessary} feature, is an open research topic. 
The current study focuses on the ground-state identification; adding temperature and entropic effects would be an interesting direction to take.

\textit{Acknowledgement.} This work was financially supported by the National Natural Science Foundation of China (12275137) and Natural Sciences and Engineering Council of Canada. We thank the Digital Research Alliance of Canada for providing computational
resources.

\textit{Author Contributions.} Y. L. designed the project. W. L. performed the MD
simulations and H. X. performed the BNT analysis. B. L., J. Z. Y. C., and Y. L. supervised the project. All authors are involved in writing and revising the manuscript.

\textit{Competing interests.} The authors declare no competing interests.

\bibliography{refs}
\clearpage

\appendix

\section{Simulation method and Sq area fraction}
\label{App:simulation}

\begin{figure}[tbp]
\centering
\includegraphics[width=\columnwidth]{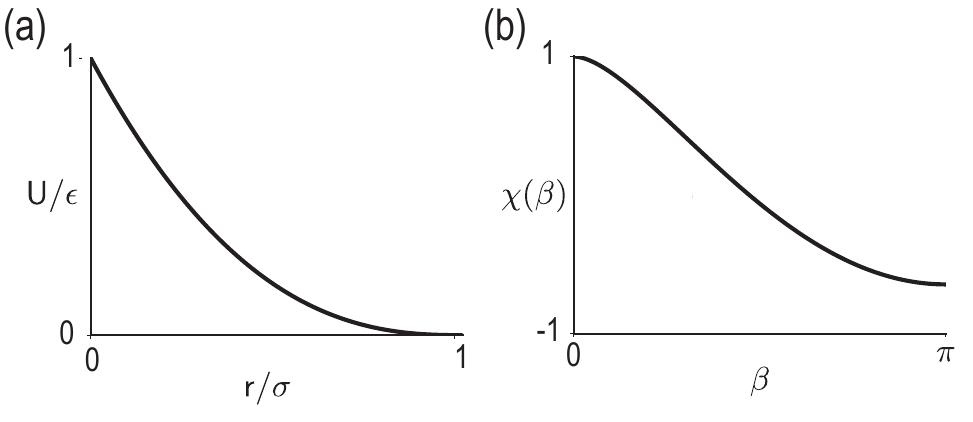}
\renewcommand{\thefigure}{A1}
\caption{
(a) Hertizan interaction $U$ in Eq. \ref{hertzian}. (b) BNT interaction $\chi(\beta)$ in Eq. \ref{BNT}
}
\label{app_fig1}
\end{figure}

The MD simulation method was used in the text for finding the near-ground-state energy of various states. 
Confined on the surface of a sphere of radius $R$ are  $N$  particles ($N=750$ to $4000$) in a single layer. 
A repulsive potential energy between particles $j$ and $k$ is assumed to follow the Hertzian form, 
\begin{equation}
U(r_{jk})=\begin{cases}
\epsilon\left(1-{r_{jk}}/{\sigma}\right)^{5/2}, & \text{if } {r_{jk}}\leq {\sigma},\\
0, & \text{otherwise},
\end{cases}
\label{hertzian}
\end{equation}
where  $r_{jk}=|{\bf r}_{j}-{\bf r}_{k}|$, and 
${\bf r}_{j}$ and ${\bf r}_{k}$ are the positional vectors of the $j$th and $k$th  particles. 
The interaction strength $\epsilon$ and force range 
$\sigma$ are separately used as the units for energy and distance.
The interaction function is plotted in Fig. \ref{app_fig1}(a). 

The simulated annealing Langevin dynamics provided by
LAMMPS is used \cite{LAMMPS}. 
 In LAMMPS, the damp constant $\gamma=\frac{m}{damp}$, where $m$ is the mass of particles, and $damp$ is the damping coefficient. In this work, $damp$ is set to $10dt$, with the time step $dt=0.5$. The time unit is $\sqrt{m\sigma^{2}/\epsilon}$. 

Initial particle configurations are randomly generated on the spherical surface 
at a specified reduced density $\rho^{\ast}=N\sigma^2/4\pi R^2$. The system is initialized at a temperature $T_{0}=10^{-2}$ in units of $\epsilon/k_B$.
The Rattle algorithm is employed to enforce the geometric constraints.
After an initial equilibration period of $10^{6}$ time steps at the
starting temperature, a simulated annealing process with $n=2500$
steps is carried out. The temperature decreases according to the relation
$T=T_{0}\exp(-0.002\times n)$, progressively cooling the system toward
a target temperature $T<10^{-4}$. 
At each temperature, $10^{4}$
time steps are performed, 
Subsequently, the system's energy 
is minimized by backtracking algorithm.

In the context of
Hex-Sq lattice coexistence, $\rho^{\ast}$ determines the position
of the crystal phase on the spinodal curve, thus dictating the relative
proportions of the Hex and Sq lattices. 
Throughout this paper, $f_{\text{Sq}}$ is used to represent the area fraction of Sq lattices compared to the overall area of the spherical surface. Directly calculating $f_{\text{Sq}}$ from the simulation results encounters certain difficulty, due to Sq lattices undergoing slight deformation and the presence of defects. Here, we estimate
the Sq area fraction using {$f_{\text{Sq}}={L_{\text {Sq}}}^2N_{\text {Sq}}/4\pi R^2$,} 
where {$N_{\text{Sq}}$ and  $L_{\text{Sq}}$ are  the number and average bond length of Sq cells in the system.}

\section{Defects and winding number}
\label{App:defect}

\begin{figure}[tbp]
\centering
\includegraphics[width=0.9\columnwidth]{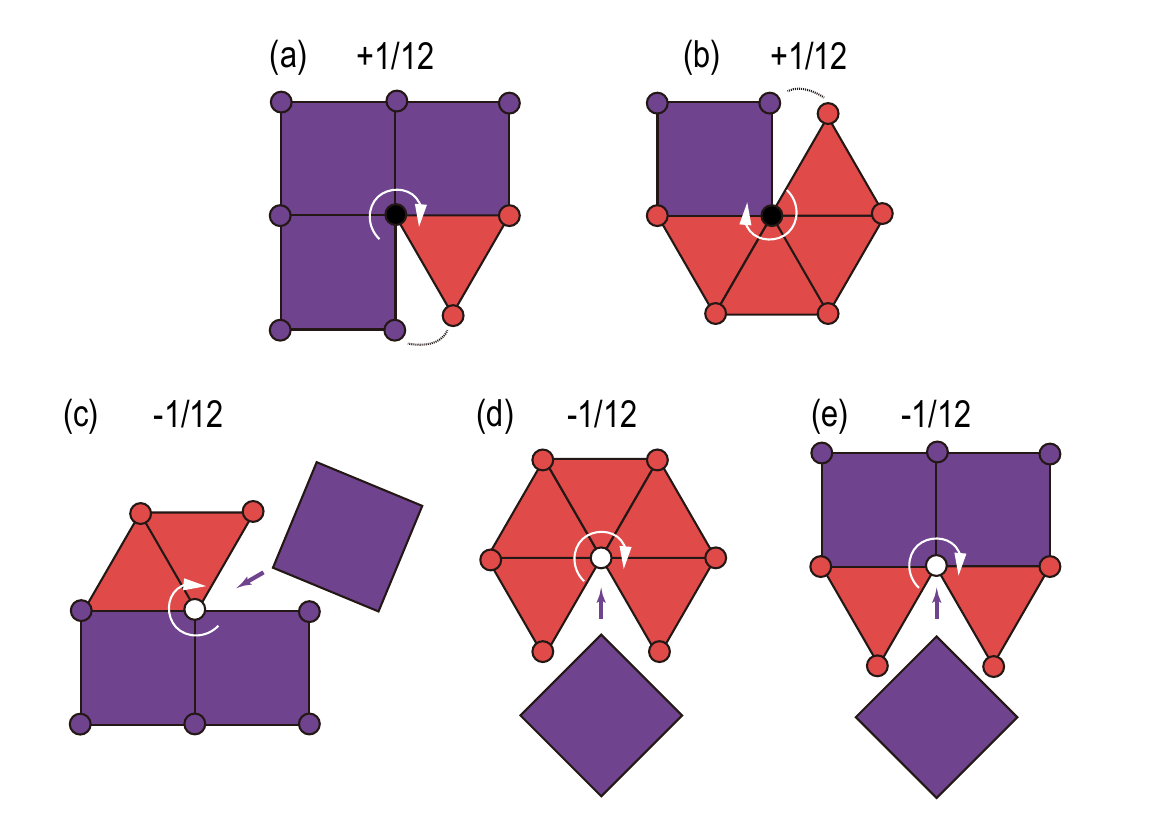}
\renewcommand{\thefigure}{A2}
\caption{Schematic illustration of winding number calculation: (a)-(b) a triangular cell replacing a square produces a $+1/12$ defect (black dots in Fig. \ref{Fig1}), (c)-(e) while a square replacing a triangle produces a $-1/12$ defect (white dots in Fig. \ref{Fig1}).}
\label{app_fig2}
\end{figure}

In the context of lattice disclinations, the winding number provides a measure of their strength by quantifying how much the local lattice orientation “rotates” when encircling the defect. In a defect-free lattice, completing a full circuit results in a total rotation of $2\pi$, bringing the orientation back to its original state. By contrast, the presence of a disclination alters this cumulative rotation, making it either less or greater than $2\pi$. The winding number is defined accordingly: if the rotation falls short of $2\pi$ by an angle $\theta$, the winding number is $w=+\theta/2\pi$, while if it exceeds $2\pi$ by $\theta$, the winding number is $w=-\theta/2\pi$. A schematic example of this calculation in the current work is illustrated in Fig.~\ref{app_fig2}. When a triangular lattice cell replaces a square, as plotted in Fig.~\ref{app_fig2}(a)-(b), a $+1/12$ defect is introduced; conversely, when a square replaces a triangle, as shown in Fig.~\ref{app_fig2}(c)-(e), a $-1/12$ defect emerges. 

In essence, the winding number captures the angular mismatch introduced by a disclination, providing a tool to classify lattice deformations around defects. Importantly, the sum of winding numbers in a system equals the Euler characteristic $\chi=V-E+F$, where $V$, $E$, and $F$ are the numbers of vertices, edges, and faces of the underlying topology \cite{poincare1895analysis}. For instance, $\chi=2$ for convex polyhedra and spherical surfaces.

\section{Fitting parameters $b_b$ and $b_{-1/12}$ as functions of $f_{Sq}$}
\label{App:fitting}

\begin{figure}[tbp]
\centering
\includegraphics[width=0.9\columnwidth]{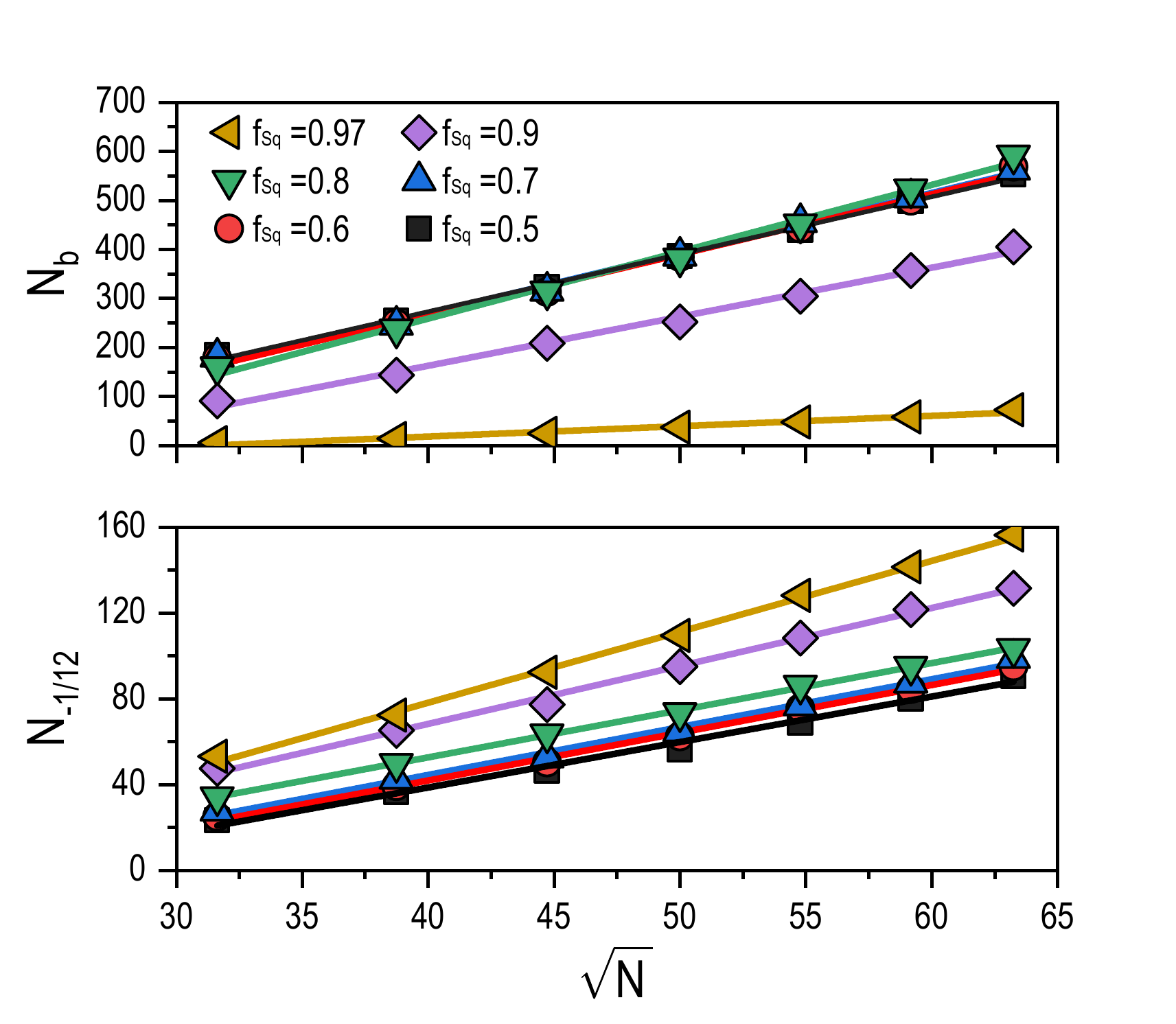}
\renewcommand{\thefigure}{A3}
\caption{The relationships of $N_b$ and $N_{-1/12}$ with $\sqrt{N}$ are fitted to the forms $N_b \sim a_b + b_b {\sqrt N}$ and $N_{-1/12} \sim a_{-1/12} + b_{-1/12} {\sqrt N}$, respectively. The fitting parameters $b_b$ and $b_{-1/12}$ are shown in Fig. \ref{FIG2}(c)-(d) as a function of $f_{Sq}$.}
\label{app_fig3}
\end{figure}

The data from Fig. \ref{FIG2}(a)-(b) were extracted and subjected to linear fitting for $N_b$ and $N_{-1/12}$, specifically in the forms
$N_b \sim a_b + b_b \sqrt{N}, 
N_{-1/12} \sim a_{-1/12} + b_{-1/12} \sqrt{N}.$
Using the ordinary least squares (OLS) method, the fitted curves are presented in Fig. \ref{app_fig3}, where the shapes colored brown, purple, green, blue, red, and black correspond to $f_{sq} = 0.97, 0.9, 0.8, 0.7, 0.6,$ and $0.5$, respectively. These figures clearly demonstrate the linear relationships of $N_b$ and $N_{-1/12}$ with $\sqrt{N}$, and the slopes of line $b_b$ and $b_{-1/12}$ are shown in Fig. \ref{FIG2}(c)-(d).

\section{BNT model}
\label{Appendix A}

To describe the global elastic energy of the entire defect pattern, the BNT potential energy is used for defect-defect interaction in the text.  It deals with a set of $n$ point disclinations on a spherical surface of radius $R$,
\begin{equation}
E_{\text{BNT}} = \frac{1}{2} R^2 \sum_{i=1}^{n} \sum_{j=1}^{n} w_i w_j K_{ij} \chi(\beta_{ij}) + \sum_{i=1}^{n} c(w_i),
\label{BNT}
\end{equation}
where the two terms correspond to the interaction energy between disclinations and the core energy required to create them \cite{Bowick2000interacting}.

The quantity $w_i$ is the winding number of the $i$-th disclination, and $K_{ij}$ is the reduced stretching modulus between disclinations $i$ and $j$. The function $\chi(\beta_{ij})$, shown in Fig.~\ref{app_fig1}(b), is a geometric kernel that depends only on the angular separation $\beta_{ij}$ between the two defects, and is defined as
\begin{equation}
\chi(\beta) = 1 + \int_{0}^{(1 - \cos\beta)/2} \frac{\ln(z)}{1 - z} \, dz.
\label{betadef}
\end{equation}
The second term, $\sum_{i=1}^{n} c(w_i)$, represents the core energy of isolated disclinations, where $c(w_i)$ is the energy cost of creating a defect with winding number $w_i$.  

To enable a direct comparison between the unbridged configuration [Fig.~\ref{Fig1}(b)] and the bridged configuration [Fig.~\ref{Fig1}(c)], the following simplifications are made. (i) Both structures are considered in the continuous limit, for which the BNT energy can be employed. 
(ii) A single-modulus approximation, $K_{ij} = K$, is used, such that interactions through different media (Hex and Sq) are treated equally. (iii) Both states contain an identical number of disclinations, $n = 24$, each carrying the same winding number $w = +1/12$.  

Under these conditions, and with the removal of the last sum in \eqref{BNT}, which is a constant, we define a
reduced energy per area. Within a numerical coefficient, 
the reduced energy $\Delta E$  
\begin{equation}
\label{DeltaE}
\Delta E = \frac{1}{2} \sum_{i=1}^{24} \sum_{j=1}^{24} \chi(\beta_{ij}),
\end{equation}
is used in \eqref{effect_BNT} for both bridged and unbridged states. 

\section{Construction of unbridged and bridged configurations}
\label{appendix B}

In the unbridged state, 24 defect points are located at the vertices of the eight equilateral spherical triangles, as illustrated in Figs. \ref{FIG3}(e) and (g).
The centers of these triangular domains {form the corners of the cube}. 
Two neighboring spherical triangles are oriented to exhibit parallel edges along their adjacent sides. 

\begin{figure}[tbp]
\centering
\includegraphics[width=0.9\columnwidth]{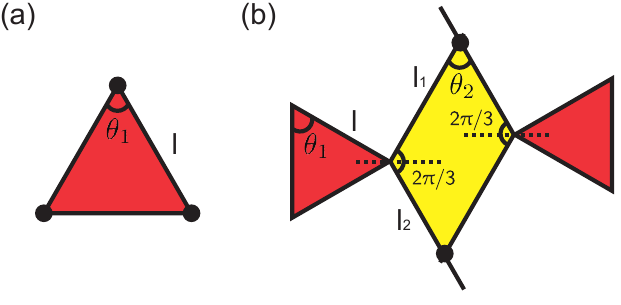}
\renewcommand{\thefigure}{A4}
\caption{
Flattened-out views to illustrate
the construction for (a) unbridged cubic configurations and (b) bridged {cubic} configurations. {The color scheme is consistent with that in Figs}. \ref{Fig1} and \ref{FIG3}}

\label{app_fig4}
\end{figure}

To calculate the total energy, 
we start by constructing eight such triangular domains on a spherical surface, which have identical edge geodesic length $l$, as shown in the flattened-out illustration in Fig. \ref{app_fig4}(a). The area of each triangular domain $S_T$ is calculated by {the} Gauss-Bonnet theorem:
\begin{equation}
S_T = R^2 (3\theta_1-\pi),
\label{S_T}
\end{equation} 
The assumption is that these triangle 
domains are Hex-tiled, thus the Sq area fraction is $f_{\text{Sq}}=1-(8S_T/4\pi R^2)$. 

For a given $f_{\text{Sq}}$, the only degree of freedom {(DOF)} is the rotation of the triangles about the spherical axis 
going through the triangle center. Minimization of the total BNT energy with respect to the rotational angle gives the black solid curve in Fig. \ref{FIG3}(b).

In the bridged state, the centers of {the} eight triangular-shaped regions are also located 
at the corners of the cube. The relative orientation of the two adjacent triangles 
is such that the triangles point to the opposite directions, in a flattened-out view [see Fig. \ref{app_fig4}(b)]. The area of these triangular domains is still given by Eq.~\ref{S_T}. 

Between the two spherical triangles, {the reverse extension of the angle bisector serves as the bisector of a new $2\pi/3$ angle, with its two sides (rays) intersecting the two rays of the adjacent triangle domain at two distinct points}, forming a bridging parallelogram, colored yellow in Fig. \ref{app_fig4}(b). This construction simulates the bridging domains observed from MD simulations shown in Fig. \ref{Fig1}(c). On the spherical surface, the parallelogram has two interior angles $\theta_2$ and two interior angles $2\pi/3$. It has 
 the area
\begin{equation}
S_P = R^2 \left(2 \times \frac{2\pi}{3} + 2\theta_2 - 2\pi\right) = R^2\left(2\theta_2-\frac{2\pi}{3}\right).
\end{equation}
Both the triangular domains (red) and parallelogram domains (yellow) are tiled with Hex units. Therefore, the fraction of Sq tiles, $f_{\text{Sq}}$.  for the entire system is 
\begin{equation}
f = 1 - \frac{8S_T + 12S_P}{4\pi R^2},
\label{C3}
\end{equation}
which is fixed once $\theta_1$ and $\theta_2$ are given.

The first DOF for bridged states is the rotation of the red-shaded triangular faces about the spherical axis passing through each triangle’s center. This face rotation alters both the shape and area of the bridging parallelogram, thereby modifying the global defect symmetries, the Sq fraction $f_{\text{Sq}}$ and $\theta_2$. 

The second DOF arises from the area redistribution between triangles and parallelograms. For a given $f_{\text{Sq}}$, the areas $S_T$ and $S_P$ can vary while satisfying the constraint in Eq. \ref{C3}; for instance, the triangles may shrink as the parallelograms expand, and vice versa. This DOF allows the structure to search for a lower energy configuration, significantly enhancing the geometric flexibility of the bridged states. Under this geometric constraint, the DOF associated with $\theta_2$ becomes dependent on $\theta_1$ and the triangle rotational angle. 

Consequently, BNT energy at a given $f_{\text{Sq}}$ in bridged states becomes a function of two parameters: the rotational angle of the triangles and $\theta_1$, thus possessing one extra DOF compared to unbridged cubic states. Minimizing the total BNT energy with respect to these two parameters yields the red solid curve shown in Fig. \ref{FIG3}(b).

\end{document}